\documentclass[11pt]{article} 
\usepackage{hyperref} 
\pdfoutput=1 
 
\begin{document} 
 
\title{Liquid Pearls} 
 
\author{
Nicolas Bremond
 \and
 Enric Santanach-Carreras
\and
J\'er\^ome Bibette \\ 
\\\vspace{6pt}  ESPCI ParisTech, LCMD, UPMC Univ. Paris 06, UMR 7195,\\ 10 rue Vauquelin, 75231 Paris, France} 
 
\maketitle 
 
 
\begin{abstract} 
This fluid dynamics video reports how to form liquid core capsules having a thin hydrogel elastic membrane named liquid pearls. These fish-egg like structures are initially made of a millimetric liquid drop, aqueous or not,  coated with an aqueous liquid film containing sodium alginate that gels once the double drop enters a calcium chloride bath. The creation of such pearls with micrometer thick membrane requires to suppress mixing until gelling takes place. Here, we show that superimposing a two dimensional surfactant precipitation at the interface confers a transient rigidity that can damp the shear induced instability at impact. Based on this, pearls containing almost any type of liquids can be created. The video focuses on the dynamics of the entry of the compound drop into the gelling bath.
\end{abstract} 
 
 
\section{Introduction} 
 
 In this video, we describe the formation of liquid pearls with millimetric scale diameter having a few micrometer thick hydrogel alginate membrane. These pearls can contain almost all type of liquids from pure water to oil. Their formation is based on  a two step procedure : the breakup of a compound pending drop followed by the solidification of the coating layer  once the compound drop enters a gelling calcium bath. Making the skin thin enough but still being able to turn into an elastic sheet is shown to be possible by controlling the interfacial properties with surfactants.  Here we show how to suppress the most important instabilities inherent to that two step procedure. Indeed, the main instability arises from the mixing of the thin alginate solution layer with the surrounding miscible liquids. This is suppressed by inducing a two dimensional surfactant solidification at the drop interface which surprisingly transforms this interface into a transient elastic membrane. This allows the persistence of the surrounding layer until gelling takes place allowing formation of few micrometer thick membrane.
 
 \section{Video information} 
 
 The compound drop is formed from two coaxial tubes in a dual dripping regime. The outer diameter $D$ of the outer needle, that sets the drop diameter, is 3 mm. The flow of both liquid phases are driven by syringe pumps and the ratio of the flow rates $r_q$ fixes the average thickness of the shell.  Depending on the outer solution surface tension and the core density, the compound drop radius $R$ varies between 1.8 mm and 2.2 mm and the flow rate ratio  $r_q$ spans from 1 to 1000 that lead to an average thickness $h$ that varies from 0.6 $\mu$m to 150 $\mu$m.
 
 The outer fluid is an aqueous solution containing 20 g of sodium alginate( a polysaccharide produced by brown algae), for 1000 g of milli-Q water.  The alginate solution exhibits a shear thinning behavior with a viscosity of 5 500 mPa.s at a zero frequency. The interfacial properties are modified by adding to the alginate solution  an anionic surfactant, sodium dodecyl sulfate (SDS) that is characterized by a critical micellar concentration (cmc) close to 8 mM. The pure alginate solution is characterized by a surface tension $\gamma_\textrm{o}$ of 72 mN/m that drops down to 35 mM/m when 10 mM of SDS is added.  Milli-Q water is used as aqueous core. For hydrophobic cores, we choose silicone oils of various viscosities that span from 1 mPa.s to 1000 mPa.s. The surface tension $\gamma_\textrm{i}$ of the oils is around 21 mN/m and the interfacial tension $\gamma_{\textrm{io}} $ between the oil and a pure alginate solution  is 40 mN/m  and drops down to 14 mM/m when 10 mM of SDS is added into the solution.\\ 
 The compound drop is then gelled once it enters a water bath containing 15 \% in weight of calcium chloride (CaCl$_2$) by diffusion of the calcium ions through the alginate solution. The CaCl$_2$ concentration leads to a density of the solution of 1130 kg/m$^3$.\\

 The impact  of the compound drop into the gelling bath are visualized with the help of a high speed camera through a macro lens. The impact dynamics is shown in the present video and helps to understand how the formation of such liquid pearls is made possible by controlling the physicochemical conditions of the fluid interfaces.
 
  \section{References} 
  
  More information can be found in the following paper:\\
  N. Bremond, E. Santanach-Carreras, L. Y. Chu, J. Bibette, 2010, \textit{Soft Matter} \textbf{6}, 2484-2488

\end{document}